# Aromaticity in a Surface Deposited Cluster: Pd$_4$ on TiO$_2$ (110)

Jin Zhang and Anastassia N. Alexandrova*

Department of Chemistry and Biochemistry, University of California, Los Angeles, Los Angeles, CA 90095-1569

Supporting Information Placeholder

**ABSTRACT:** We report the presence of σ-aromaticity in a surface deposited cluster, Pd$_4$ on TiO$_2$ (110). In the gas phase, Pd$_4$ adopts a tetrahedral structure. However, surface binding promotes a flat, σ-aromatic cluster. This is the first time aromaticity is found in surface deposited clusters. Systems of this type emerge as a promising class of catalyst, and so realization of aromaticity in them may help to rationalize their reactivity and catalytic properties, as a function of cluster size and composition.

Surface-deposited clusters composed of only a few atoms gain popularity as a new class of potent heterogeneous catalysts.[1-17] Small clusters are even suspected to be responsible for the majority of the catalytic activity, when most of the material is present in the form of much larger particles.[12] For better or for worse, chemical properties of small clusters are exceptionally sensitive to cluster size, and so their use in catalysis critically depends on our understanding of chemical bonding in them, and thus the ability to rationally exploit their electronic properties. However, no qualitative theory of chemical bonding that normally guides chemists in predicting properties of a chemical system is developed for surface deposited clusters, which therefore remain a chemical enigma.

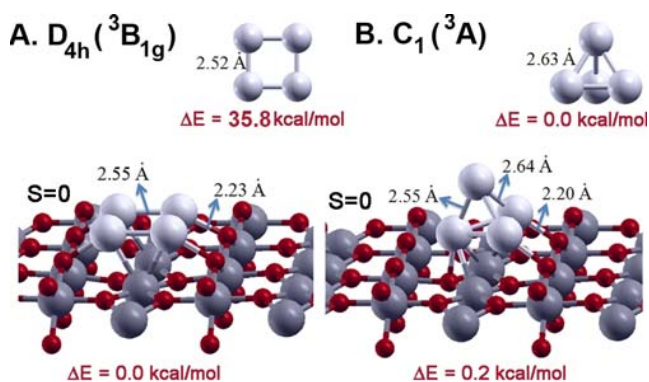

Figure 1. The two most stable isomers of Pd$_4$ in the gas phase and on the TiO$_2$ (110) stoichiometric surface. In the gas phase, the Jahn-Teller distorted tetrahedral species is the global minimum. However, on the surface, the square species gets significantly stabilized, and becomes more stable than the tetrahedron. In both cases, the open spins are quenched upon surface binding. Geometric parameters are shown to illustrate the strain imposed on the equilibrium structures by surface binding. All results are obtained with the Density Functional Theory, PBE, with the plane wave basis set.

Among systems of a particular interest are clusters of Pd deposited on the rutile, TiO$_2$, (110) surface. They were shown to be active toward CO oxidation, but only at certain sizes, and activity would inexplicably drop at other sizes.[13] For example, activity was miniscule at Pd$_1$, elevated at Pd$_2$, reduced at Pd$_4$, dropped at Pd$_7$, maximal at Pd$_{20}$, and reduced again at Pd$_{25}$. So far, this trend remains baffling, and that brings to light the lack of any qualitative chemical bonding concepts that would enable our intuition about properties of such clusters.

The cluster under the present scrutiny is Pd$_4$. We found that in the gas phase, it adopts a tetrahedral structure slightly Jahn-Teller distorted to C$_1$ ($^3$A) symmetry, with its corresponding singlet being at least 7 kcal/mol less stable. The second type of structure is square. The square isomers are significantly higher in energy than the tetrahedral species, and the square triplet, D$_{4h}$ ($^3$B$_{1g}$), is again more stable than the singlet, D$_{4h}$ ($^1$A$_{1g}$). This result is reproduced at a variety of levels of theory, including MP2 and CCSD(T), with different basis sets (Supporting Information).

However, when Pd$_4$ is deposited on the stoicheometric TiO$_2$ (110) surface, the situation changes. The flat square structure becomes the global minimum (Figure 1A). The unpaired spins are quenched, so that the system becomes a singlet. The singlet tetrahedral structure becomes higher in energy, though only slightly (Figure 1B). All small Pd$_n$ ($n$=1-4) clusters preferentially bind to the stoicheometric surface, and repel oxygen vacancies (the most typical defects in rutile).[18] Therefore, only the stoicheometric surface is considered here.

The square structure being preferred for the surface bound cluster is foremost attributable to the better matching of the cluster with the underlying crystal lattice. Both types of clusters exhibit a partially covalent binding to the surface, and the states of the cluster mix mostly with those of the surface oxygen atoms. For the square species, this binding to oxygens is more geometrically feasible. As a result of this binding, partial charge transfer also happens from the clusters to the surface: 0.44 electrons in the case of the square, and 0.32 electrons in the case of the tetrahedron, again indicating stronger binding in the case of the square. When the tetrahedron is deposited on the surface, it has to undergo a larger structural distortion from its equilibrium gas phase geometry (illustrated in Figure 1), experiencing strain and destabilization, greater than those for the square structure. However, more importantly its binding causes a costly reconstruction of the underlying surface. The energy penalty for this reconstruction is ca. 30 kcal/mol greater for the tetrahedral cluster than for the square one (see Table S2, Supporting Information). Even though the adsorption energy for the tetrahedral Pd$_4$ is only 2.3 kcal/mol lower than that for the square structure, the penalty for distortion makes the tetrahedron disfavored.

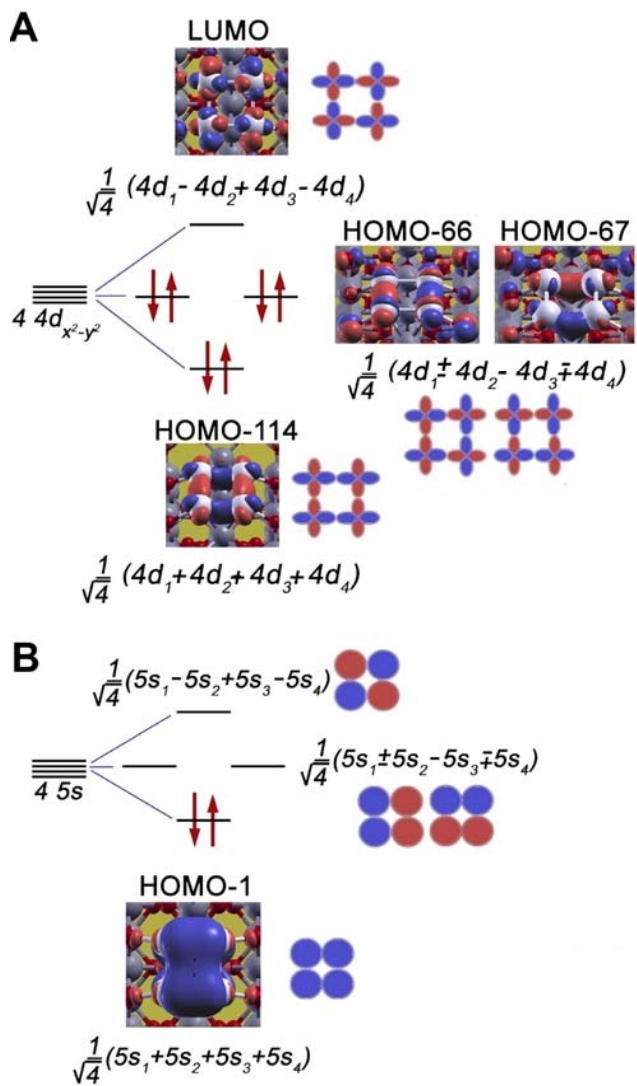

Figure 2. MOs contributing to the intracluster bonding in the square $Pd_4$ on $TiO_2(110)$: (A) The four $4dx^2$-$y^2$ AOs overlap in the cluster plane, and also form four σ-MOs. One of the four MOs resulting from the 4 $4dx^2$-$y^2$ AOs is unoccupied, and is the LUMO. The set is, therefore, populated by 6 electrons, which gives the cluster the second type of σ-aromaticity. No other 4d-AOs contribute to bonding significantly. (B) The HOMO-1 formed by 5s-AOs on Pd atoms. The inset illustrates how the four 5s-AOs form four MOs: one bonding, a non-bonding pair, and one antibonding. Only the bonding MO is populated in the cluster. The two electrons in this MO make the cluster σ-aromatic. Schematic representations are shown under each of the MOs for clarity.

Furthermore, the surface promotes $Pd_4$ going both planar, and σ-aromatic, as follows. The electronic configuration of the Pd atom is $[Kr]d^{10}$. The 5 4d atomic orbitals (AOs) on each Pd atom together yield a total of 20 molecular orbitals (MOs) in the cluster, i.e. 5 sets of the bonding, doubly-degenerate nonbonding, and antibonding MOs. If all these MOs would be fully populated, and no other valence MOs would be present, the net bonding in the cluster would be zero, and $Pd_4$ would be unbound. However, only 19 of these MOs are occupied in the singlet square species. One of them, the antibonding MO formed by the $4dx^2$-$y^2$–AOs on Pd atoms, is empty in the singlet square species, and is the LUMO (Figure 2A).

The electron density removal from the antibonding MO formed by the $4dx^2$-$y^2$–AOs contributes to cluster binding. The occupied three MOs formed by $4dx^2$-$y^2$–AOs are not localizable, as for example lone pairs on atoms, or 2 center – 2 electron bonds. They overlap in the cluster plane, in the σ-fashion, i.e. in every pair of overlapping $4dx^2$-$y^2$–AOs each AO uses only one lobe in the overlap. Thus, the considered subsystem of MOs in the $Pd_4$ cluster is populated by 6 delocalized σ-electrons. This renders the system σ-aromatic, according to the $(4n+2)$ Hückel's rule for aromatic compounds, with $n=1$.

The rest of the 4d-AOs ($dxy$, $dyz$, $dxz$, $dz^2$) form four completely populated sets of four MOs each, including bonding, nonbonding, and antibonding. Therefore, their net bonding effect in the cluster is nearly zero, and these MOs can be thought of as lone pairs on Pd atoms.

The electron spill happens from the 4d-AOs (the LUMO) to the 5s-AOs in $Pd_4$, and this also contributes to cluster binding. Apparently, it is energetically favorable to recruit the higher energy 5s-AOs into the set of valence MOs. Four 5s-AOs, slightly mixed with the $4dz^2$-AOs having the same symmetry, form four linear combinations (Figure 2B). Only one of these MOs, the bonding HOMO-1, is populated in the cluster. It is a completely symmetric delocalized MO of a σ-type, and it contributes to the σ-bonding in the system. Populated by 2 electrons, the HOMO-1 makes the cluster again obey the $(4n+2)$ Hückel's rule, with $n=0$ this time. Hence, the cluster is σ-aromatic due to this set of MOs as well. Therefore, the surface-bound singlet square cluster is doubly σ-aromatic. Again, without double σ-aromaticity, the singlet square $Pd_4$ cluster would be unbound, both in the gas phase and on the surface.

The chemical bonding in the square structure in the gas phase is very similar to that on the surface. The gas phase cluster is also doubly aromatic (see Supporting Information). However, the square cluster in the gas phase is a metastable minimum, which would be most likely not observable experimentally. Therefore, aromaticity in $Pd_4$ is surface-binding promoted. The surface deposition does not make $Pd_4$ σ-aromatic, but stabilizes the flat cluster, and thus allows for the expression of double σ-aromaticity in it. σ-aromaticity has been previously observed in gas phase metallic clusters.[19-39] However, this is the first time that it has been found in a surface deposited cluster.

How is the tetrahedral $Pd_4$ structure bound? Both in the gas phase, and on the surface, the singlet cluster possesses 19 occupied MOs formed by 4d-AOs on atoms. The displaced d-electron density goes to the completely bonding σ-MO formed by 5s-AOs, just like in the square structure. One of the differences between the tetrahedron and the square is that in the former, the 5s-AOs overlap in 3-D, instead of 2-D. The 3-D overlap is more efficient for 5s-AOs, and so the tetrahedral structure is more stable than the square in the gas phase. On the surface, the tetrahedral species experiences a greater strain and distortion, as was discussed, and so it is destabilized.

Finally, we would like to recall that, in the dependence of the catalytic activity of rutile-deposited $Pd_n$ on cluster size, $Pd_4$ showed a reduced catalytic activity. It is possible that the activity drop at $Pd_4$ has to do with its enhanced stabilization, in part owing to its σ-aromatic character. Aromaticity is a chemical bonding phenomenon generally associated with relative stability, and a tendency to undergo reactions of substitution rather than addition. In the catalytic process, the binding of substrates (CO and $O_2$) would necessarily disturb the resonance and delocalized aromatic bonding in $Pd_4$. Therefore, from this prospective, binding the substrates and transition states of the reaction should be disfavored. This, however, remains to be demonstrated.

## ASSOCIATED CONTENT

**Supporting Information**. The description of computational methods, the relative energies of isomers in the gas phase obtained with different computational methods, the energy penalties due to cluster and surface distortion upon binding, and full MO diagrams for the gas phase and surface deposited clusters. This material is available free of charge via the Internet at http://pubs.acs.org.

## AUTHOR INFORMATION

### Corresponding Author

ana@chem.ucla.edu

## ACKNOWLEDGMENT

Financial support for this work was provided through the ACS PRF grant 51052-DNI6. Most calculations were performed on the UCLA Hoffman2 shared cluster. A portion of research was performed using the EMSL, a national scientific user facility sponsored by the Department of Energy's Office of Biological and Environmental Research and located at Pacific Northwest National Laboratory.

## ABBREVIATIONS

AO – Atomic Orbital
MO - Molecular Orbital

SYNOPSIS TOC (Word Style "SN_Synopsis_TOC"). If you are submitting your paper to a journal that requires a synopsis graphic and/or synopsis paragraph, see the Instructions for Authors on the journal's homepage for a description of what needs to be provided and for the size requirements of the artwork.

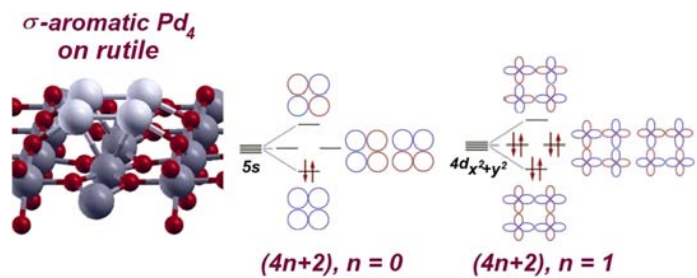